\def\aa{{A\&A}}

\def\aj{{AJ}}

\def\apj{{ApJ}}
\def\apjs{{ApJS}}

\def\mnras{{MNRAS}}

\documentclass[cup5b]{caps}
\usepackage{graphicx,psfig}
\usepackage{amssymb}
\usepackage{ociwsymp1}
\HeadText{K. Gebhardt}

\begin{document}

\pagenumbering{arabic}

\author[]{KARL GEBHARDT\\The University of Texas at Austin}

\chapter{Influence of Black Holes on \\ Stellar Orbits}

\begin{abstract}

We review the current state of dynamical modeling for galaxies in
terms of being able to measure both the central black hole mass and
stellar orbital structure. Both of these must be known adequately to
measure either property. The current set of dynamical models do
provide accurate estimates of the black hole mass {\it and}\ the
stellar orbital distribution. Generally, these models are able to
measure the black hole mass to about 20\%--30\% accuracy given present
observations, and the stellar orbital structure to about 20\% accuracy
in the radial to tangential dispersions. The stellar orbital structure
of the stars near the galaxy center show strong tangential velocity
anisotropy for most galaxies studied. Theoretical models that best
match this trend are black hole binary/merger models. There is also a
strong correlation between black hole mass and the contribution of
radial motion at large radii. This correlation may be an important
aspect of galaxy evolution.

\end{abstract}

\section{Introduction}

The first observational evidence that black holes are common in the
centers of nearby galaxies is reviewed in Kormendy (1993) and Kormendy
\& Richstone (1995). The initial studies concentrate mainly on
measuring the black hole mass and only somewhat included the effects
of different orbital structure. However, it was always apparent that
the assumed form for the distribution function has a considerable
effect on the measured black hole mass. Thus, the believability in the
existence of a central black hole closely paralleled the development
of more sophisticated modeling techniques that were designed to be as
general as possible.

There are two main aspects for making a general dynamical model. These
are the dimensionality of the potential and that of the velocity
ellipsoid. For the potential, we know that we have to at least model
galaxies as axisymmetric, and, for some, triaxial structure is
required (e.g., those with counterrotating cores, polar rings,
etc.). While it is important to allow the most freedom for a dynamical
model, there is a level of detail that need not be studied (at least
at present). For example, we know that no galaxy is exactly symmetric
along any axis.  Therefore, in order to provide an adequate
representation in that case, one cannot use symmetric dynamical
models, but instead must rely on $N$-body simulations --- similar to
what is done when modeling merging systems (Barnes \& Hernquist
1992). Using an $N$-body system to model each galaxy is currently not
practical, and, furthermore, may not even provide a better
understanding of the underlying physics due to the huge parameter
space inherent in $N$-body simulations. Thus, the most to gain lies in
using general dynamical models that may not accurately represent the
galaxy, but serve to provide overall trends from which we can infer
formational and evolutionary scenarios. In other words, we will always
be making some error --- no matter which dynamical models we use ---
but we should be aware of each model's limitations. Below, we first
review the dynamical models that have been applied to nearby galaxies,
and then summarize the current state-of-the-art and the overall
results from these models.


\begin{figure*}[t]
\centerline{\psfig{file=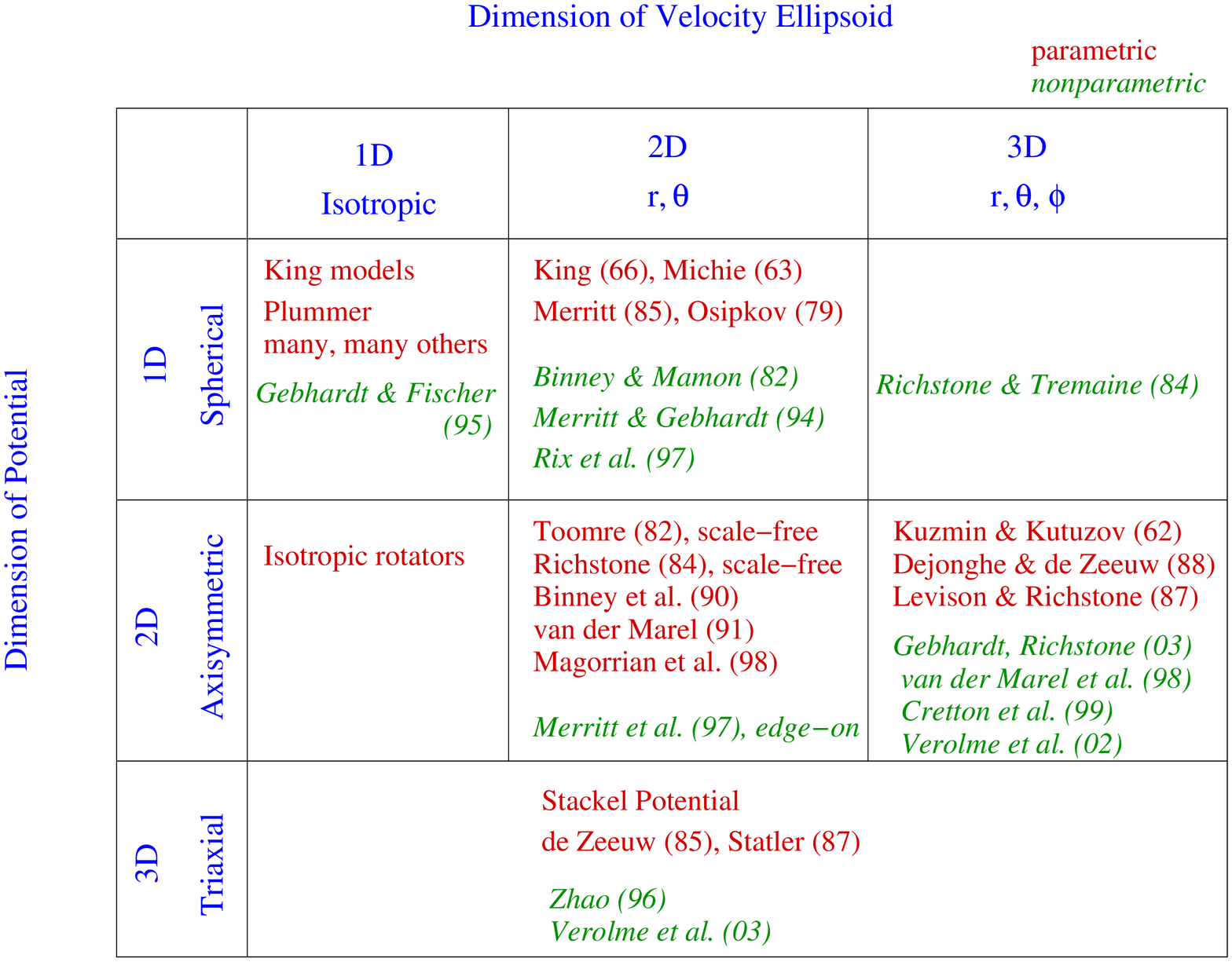,width=15cm,angle=-90}}
\caption{The two main assumptions made in dynamical models: the
dimension of the potential is along the vertical axis and that of the
velocity ellipsoid is along the horizontal axis. Each box includes a
few relevant papers for each configuration. These references are not
complete and only serve to provide examples. The text type refers to
whether the dynamical models assume a parametric (regular) or
nonparametric (italics) form for the distribution function.
\label{box}}
\end{figure*}


\section{The Suite of Dynamical Models}

Figure~1.1 diagrams the possible dynamical models based on their
complexity. The components are the number of symmetry axes for the
potential and for the velocity ellipsoid. The potential shapes clearly
represent spherical, axisymmetric, and triaxial shapes.  The velocity
ellipsoid shapes represent isotropic (distribution functions that
depend on only one integral of the motion, namely energy), 2-integral,
and 3-integral distribution functions. This plot provides the range of
possible distribution functions that can be used for dynamical
modeling where the system obeys some symmetry axes. Obvious omissions
are those systems that obey no symmetry axes.

The goal of the dynamical modeling is to determine the underlying
potential of the system as well as the orbital structure. The concern
is that, by not using a model that adequately represents the system,
the results may be significantly biased. The best way to test for
these biases is to model systems with a variety of assumptions and
compare the results.

In each grid element are examples from the literature that represent
that particular model. This listing is done to provide a few examples
each and is not intended to be complete in any way. In fact, a
complete listing would take the whole of this proceeding (but see
Binney \& Tremaine 1987 for a complete discussion). There are two
types for the text in each grid: regular text represent analytic
models, and italicized text represents nonparametric models. For
example, isotropic, spherical models (the upper left grid) encompass
an infinite number of density-potential pairs, and only King and
Plummer models are listed.  Gebhardt \& Fischer (1995) present a
nonparametric, isotropic, spherical model that determines the
potential directly.  Isotropic, spherical models have been enormously
successful in describing stellar systems, especially for globular
clusters (King 1966). For measuring black hole masses, they have done
remarkably well; for example, Kormendy (2003) shows the change in the
estimated black hole mass for M32 varies little over 20 years of data
and a range of model sophistication.  However, we are at a level now
where the quality of the data is so high that we must use the most
general models possible. Furthermore, in order to study the orbital
structure one must use nonparametric techniques; otherwise, one
restricts the form of the distribution function.


\begin{figure*}[t]
\centerline{\psfig{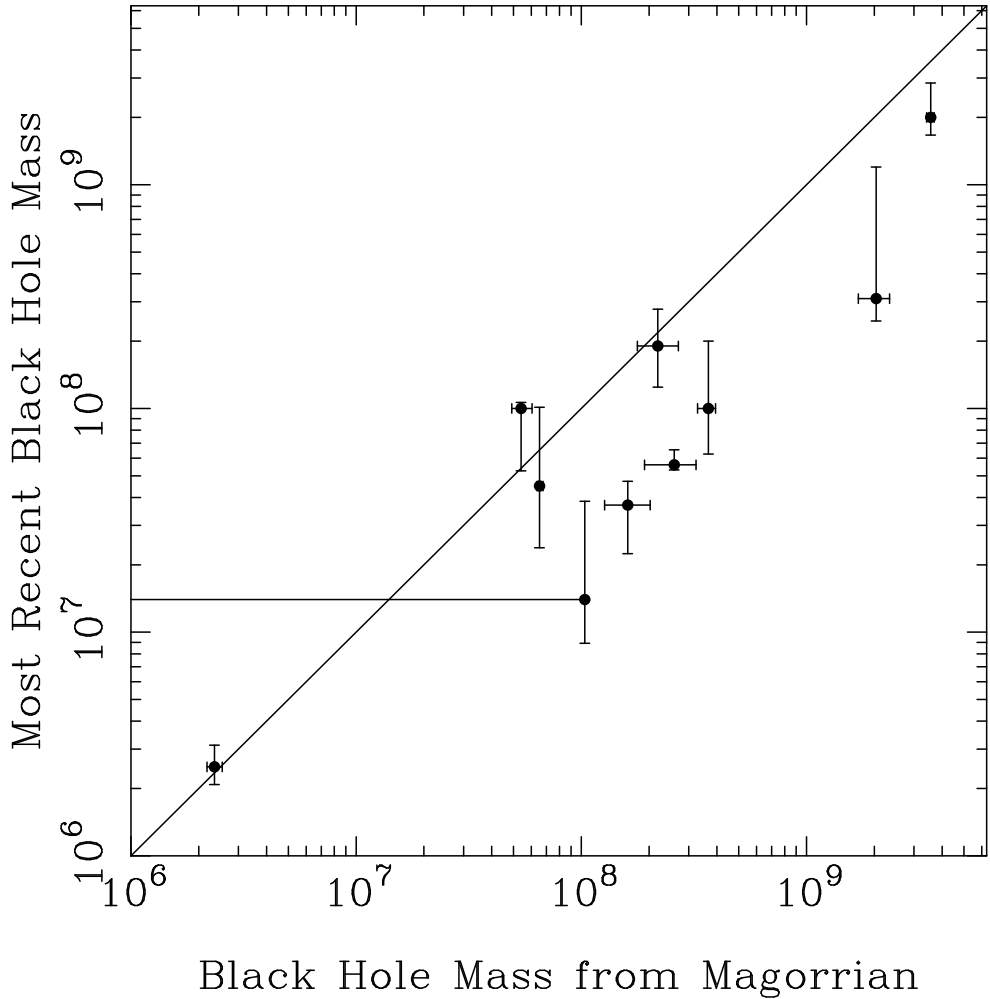}}
\vskip 0pt \caption{The black hole mass estimate from Magorrian et
al. (1998) and those from more recent data and analysis (see Tremaine
et al. 2002 for the compilation). The line is one-to-one
correspondence. All masses have been corrected to a common distance.
The Magorrian et al. (1998) masses are on average 2.4 times larger.
\label{magor}}
\end{figure*}


Spherical models are good representations for globular clusters and
some of the largest ellipticals (e.g., M87), but we know that most
galaxies are not spherical. Tremblay \& Merritt (1995) and Khairul
Alam \& Ryden (2002) argue, based on inversion of the distribution of
projected shapes, that, in fact, there are nearly no galaxies that are
spherical.  We must use, at the least, axisymmetric
models. Furthermore, Binney (1978) and Davies et al. (1983) point out
that the flattening in galaxies is not consistent with isotropic
orbits: i.e., we must also include anisotropy. Thus, there have been a
tremendous amount of work in modeling galaxies as 2-integral
axisymmetric systems.

Van~der~Marel (1991) provides one of the first 2-integral studies of a
large sample of ellipticals, using the modeling first introduced by
Binney, Davies, \& Illingworth (1990). From kinematic data taken along
the major and minor axis for 37 galaxies, van~der~Marel finds that
2-integral models have too much motion on the major axis compared to
what is seen. The implication is that ellipticals have
$\sigma_r>\sigma_\theta$, inconsistent with the 2-integral assumption
(where $\sigma_r=\sigma_\theta$). There are multiple ways to cause
this inconsistency.  For example, galaxies may depend on a third
integral of motion; the 2-integral models may be biased by not
including a dark halo; galaxies may have significant triaxial shape
which also biases axisymmetric models; or the quality of the data may
be too poor. We can compare the results of van~der~Marel to those of
Gebhardt et al. (2003), who make 3-integral models for 12 galaxies,
including three in common. For half of the sample,
$\sigma_r>\sigma_\theta$, consistent with van~der~Marel. For three
galaxies in common (NGC~3379, NGC~4649, and NGC~4697), there is not
good agreement. Neither model includes a contribution from a dark
halo, which may bias the large radial orbital structure. However, most
likely, the differences are due to the use of different data sets, and
the quality of the data sets can have a significant influence on the
results. We now turn to measuring the black hole mass.

The largest sample using 2-integral models to measure black hole
masses is that of Magorrian et al. (1998). Magorrian et al. study 36
galaxies with ground-based kinematics and {\it HST}\ photometry to
provide a systematic estimate of the central black hole mass. Previous
black hole studies concentrated on individual cases. These result have
been widely used, and also criticized. The major complaint is that the
models are still too simplistic (i.e., 2-integral) and that the
kinematic data have too low spatial resolution to say anything about
the central black hole. Many of the Magorrian et al. galaxies now have
{\it HST}\ kinematic data and have been modeled with more general
models. In Figure~1.2 we compare the black hole mass estimates from
Magorrian et al. to these more recent studies.  There is a bias in
that the 2-integral masses tend to be higher than those from the more
recent analysis. The average difference between the two samples is a
factor of 2.4. As discussed in Gebhardt et al. (2003), the difference
appears to be due to differences in modeling, as opposed to the
improved spatial resolution in the kinematics. Clearly, the better
kinematics provide a more accurate measurement, but they do not appear
to bias the results. Gebhardt et al. (2003) show that the black hole
mass is not biased when using only ground-based data compared to using
both ground-based and {\it HST}\ kinematics.

In order to provide a more accurate estimate of either the black hole
mass or the orbital structure, we need to go beyond 2-integral models.
Models that allow for three integrals of motion have only recently
been applied to dynamical systems. The problem is that the most
general form for the third integral is not analytic, and we must rely
on numerical approaches. In limiting cases, there are analytic
3-integral models; for example, Dejonghe \& de~Zeeuw (1988) study
3-integral Kuzmin-Kutuzov (1962) models. However, these models have
analytic cores ($d{\rm log}\nu/d{\rm log}r=0$ at the center), and
since nearly all galaxies have central cusps (Gebhardt et al. 1996;
Ravindranath et al. 2001), they will be of limited use. Because the
third integral is not analytic, we generally rely on orbit-based,
Schwarzschild (1979) codes in order to study them.  The first general
application of the orbit-based methods is presented in Richstone \&
Tremaine (1984), applied to spherical systems. They even incorporate
rotation in their models to provide one of the first models that
include three integrals (energy, $E$, total angular momentum, $L^2$,
and angular momentum about the pole, $L_z$), albeit in a spherical
system. Rix et al. (1997) extend this analysis to make a detailed
orbit-based model of the dark halo around NGC~2434. The first
application of an axisymmetric, orbit-based model is that of
van~der~Marel et al. (1998), who measure the black hole mass in M32. A
few groups now have axisymmetric, orbit-based codes that have been
used to study central black holes. To date, 17 galaxies have been
studied with these models, with 14 coming from one code (Gebhardt et
al. 2000, 2003; Bower et al. 2001), four from the Leiden group with
various codes (van~der~Marel et al. 1998; Cretton \& van~den~Bosch
1999; Cappellari et al. 2002; Verolme et al. 2002), and one from
Emsellem, Dejonghe, \& Bacon (1999).


\begin{figure*}[t]
\centerline{\psfig{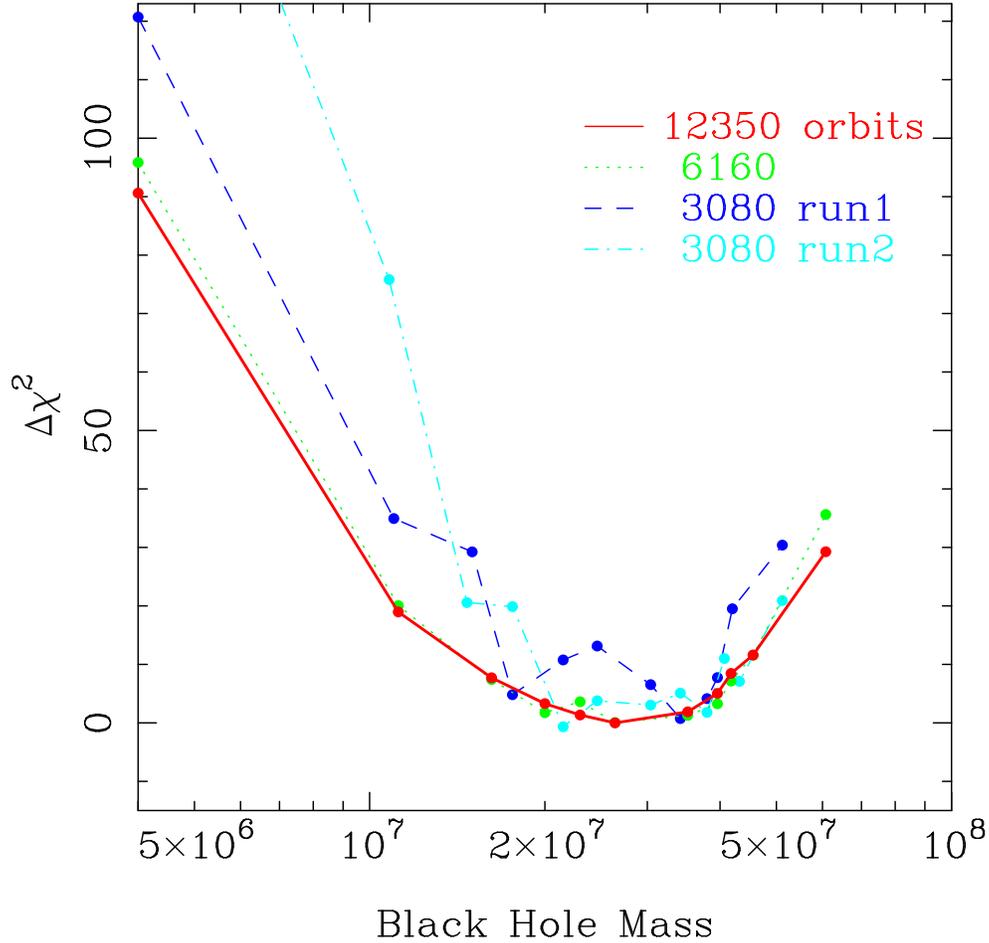}}
\vskip 0pt \caption{Shape of $\Delta\chi^2$ versus black hole mass for
models with different orbit numbers. Each model is a fit to an
identical data set, and the only difference is the sampling of phase
space. For the two runs with the smallest orbit library, we have run
the same number of orbits, but simply sampled phase space differently.
\label{norbit}}
\end{figure*}


With so few groups using orbit-based codes, we must be certain that
the immense freedom allowed by these codes does not bias the results
due to some feature of an individual code. The general problem of
covering phase space appropriately in these orbit-based codes is
tricky. There is a balance that one must obtain between including a
large orbit library in order to sample phase space but still maintain
a small enough library in order to use a reasonable amount of computer
resources.  In fact, Valluri, Merritt, \& Emsellem (2003) find that
there is a large difference when running models using orbit libraries
of various size. They have two main results that question the
reliability of these models for measuring black hole masses. First,
the shape of the $\chi^2$ contours depends on the number of orbits run
for a model, using the same data set. Second, for models with large
numbers of orbits, there is a degeneracy in black hole mass: the
$\chi^2$ contours reach a plateau over a large range of black hole
masses. These results are critically important to understand since
they may undermine this whole area of study. Fortunately, the other
groups involved have done many tests in regards to this degeneracy. We
will concentrate on the tests done with the Gebhardt et al. (2003)
code.

There are three issues on which we will focus. These are (1) the shape
of $\chi^2$ as a function of orbit number, (2) the ability of using
the $\chi^2$ contours to measure reliable confidence bands, and (3)
the dependence on the smoothing parameter. For this last aspect, most
groups use regularization for the smoothing while Gebhardt et al.
rely on maximizing entropy (Richstone \& Tremaine
1988). Regularization imposes smoothing directly in phase space by
including a term that represents the noise in the $\chi^2$, typically
using the sum of the squared second derivative between phase space
elements. Gebhardt et al. (2003) calculate the entropy of each orbit
(using entropy equal to $w{\rm log}w$, where $w$ is the orbital
weight) and use the total entropy as a constraint (see Richstone \&
Tremaine 1984, 1988 for a complete discussion). Both approaches should
provide smooth distribution functions, and there is no obvious desire
to use one over the other.


\begin{figure*}[t]
\centerline{\psfig{file=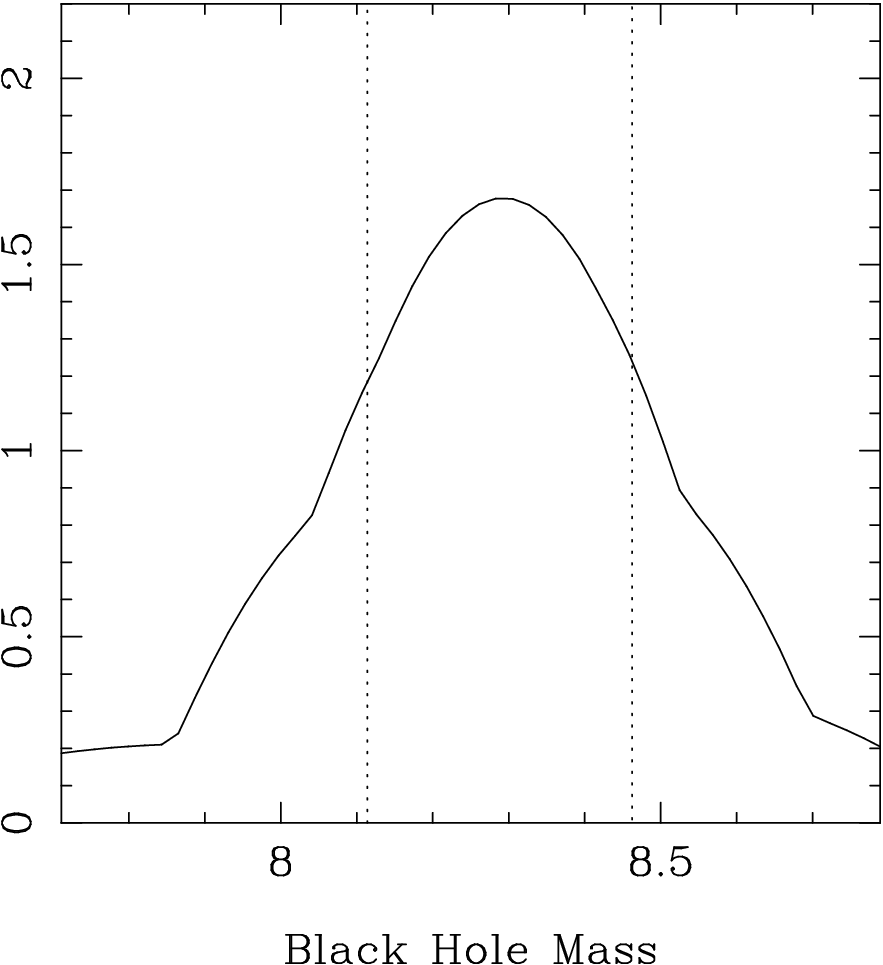,width=13cm,angle=0}}
\vskip 0pt \caption{Distribution of black hole masses from Monte Carlo
simulations for NGC~3608. The solid line represents the results of
changing the input velocity profiles according to the noise in the
spectral data (the Monte Carlo approach). The vertical dotted lines
represent the 68\% confidence limit as measured from the shape of the
$\chi^2$ contours. The area inside the dotted lines is close to 68\%
of the area.
\label{n3608bh}}
\end{figure*}


To study the influence of orbit number on the best-fit solution, the
obvious test is to run an analytic model where the black hole mass is
known and simply increase the orbit number. Valluri et al. (2003) have
the only paper in which this test has been published. This test,
however, has been done by the other groups, but it was never published
since nothing was ever seen to be problematic. Figure~1.3 plots this
test using the code of Gebhardt et al. (2003). They run four models
for the same data set. The total orbit number spans a factor of 4,
with the two smallest libraries being run twice but with a different
sampling. The two largest libraries show nearly identical $\chi^2$
profiles. The two smallest libraries show a different contour shape,
but they have substantial noise, making the comparison difficult. For
libraries with an extremely small number of orbits, it is clear that
the $\chi^2$ contours must become very noisy since the quality of the
fit depends on whether one happens to hit important orbits or
not. Thus, having an appropriate number of orbits certainly is
important. However, since we see little difference between the two
largest libraries that differ by a factor of 2 in orbit number, it
appears that the contours do not plateau as a function of black hole
mass, as Valluri et al. (2003) find. In fact, even for the small
libraries, we see that they tend to trace the true $\chi^2$ contour
fairly well, although the noise makes it difficult to follow. The
number of orbits in a given library is only useful if one compares it
to the number of model grid elements. For published orbit-based
models, most have phase space coverage that is adequate to measure the
black hole mass. For example, Gebhardt et al. (2003) use about 8000
orbits in each galaxy model with the same number of grid elements
shown in Figure.~1.3.

Another issue to understand is whether the uncertainties on the black
hole masses are adequately measured. One of the goals of black hole
studies is to understand their role in galaxy evolution, and any
comparison with galaxy properties must contain accurate uncertainties.
All orbit-based models rely on using the shape of the $\chi^2$ to
determine their uncertainties. The best method, however, is to run
bootstrap simulations on the real data. We have done this for
NGC~3608. For each spectrum, we simulate a new realization based on
the noise in the spectrum. We then generate 100 realizations. This
Monte Carlo method is the same as that used when measuring the
uncertainties for the velocity profiles (see Pinkney et al. 2003). We
then run the modeling code on each new set of data and estimate the
best-fit black hole mass.  This procedure is extremely time consuming,
and we have only done it for one galaxy so far. Figure.~1.4 plots the
distribution of black hole masses obtained by these Monte Carlo
simulations. The solid line represents the distribution function using
an adaptive kernel estimate of the individual realizations. The dotted
lines show the 68\% confidence band measured from the shape of the
$\chi^2$ contours.  The agreement is excellent, as the 68\% $\chi^2$
contours are similar to the area that contains 68\% of the
simulations. The simulations encompass a slightly larger area, but
only by a few percent. Thus, it appears that the $\chi^2$ contours can
be used to estimate accurately the black hole mass uncertainties. From
all of the orbit-based models used to date, the range of black hole
mass uncertainties is from 5\% to 70\%, with an average uncertainty
around 20\%. Given that the scatter in the $M_\bullet-\sigma$
correlation is less than 30\% (Tremaine et al. 2002), we still need to
improve the black hole mass uncertainties.


\begin{figure*}[t]
\centerline{\psfig{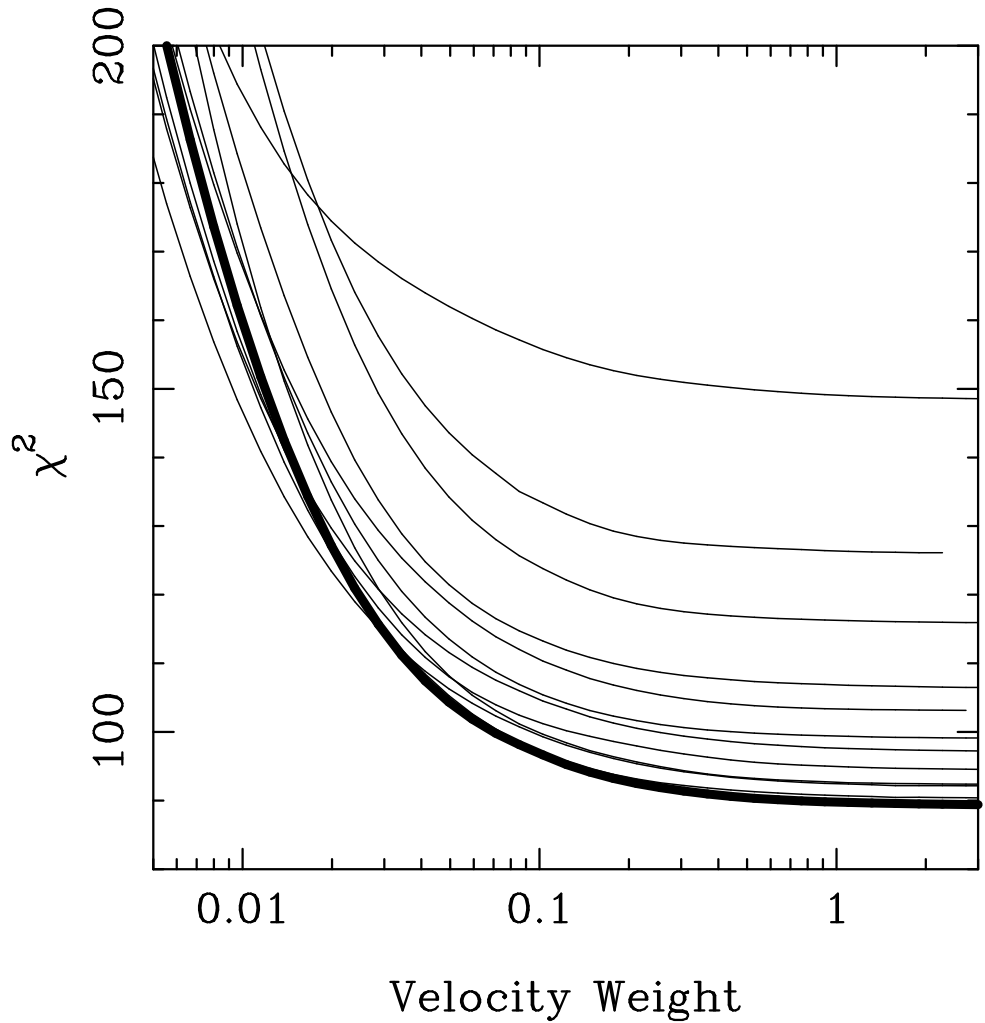}}
\vskip 0pt \caption{$\chi^2$ between the model and data as a function
of the relative weight between entropy and velocity fit for NGC~3608.
By increasing the velocity weight, we are decreasing the amount of
smoothing. We start each model with maximum smoothing (maximizing
entropy only) and then increase the velocity weight until the
kinematics are fit as well as possible. The heavy, solid line is the
best-fit model when the entropy term has no weight. The best-fit model
has the minimum $\chi^2$ over a large range of entropy weights.
\label{n3608iter}}
\end{figure*}


Another important concern is whether the smoothing parameter has an
effect on the black hole mass. The choice of this parameter is
discussed extensively in Cretton et al. (1999) and Verolme et
al. (2002).  Their choice of the smoothing parameter is based on
comparison with analytic test cases, by finding that smoothing
parameter that provides the best match for the phase space
distribution function.  This cross-validation technique is a standard
statistical approach to determine the smoothing
parameter. Furthermore, Verolme et al. (2002) have performed tests in
which they compare their best-fit mass found with optimal smoothing to
that measured when including no smoothing, and find no difference in
their black hole mass.  Similar results are found in the modeling of
Gebhardt et al. (2003).  Figure.~1.5 is a plot of $\chi^2$ versus
smoothing parameter for many different models of NGC~3608. Each line
differs by the mass of the black hole. The final $\chi^2$ versus black
hole mass is then obtained by taking the rightmost values in
Figure.~1.5. The point of this plot is to show that the best-fit model
provides the minimum $\chi^2$ over a large range of smoothing
parameters. For the maximum entropy method, the smoothness is employed
by increasing the contribution of the entropy term relative to the
comparison with the velocities. The velocity weighting is increased
until the model provides the best fit to the data and essentially
there is no contribution from the entropy term. However, as is seen in
Figure.~1.5, the best-fit model provides the minimum $\chi^2$ for a
range of 100 in smoothing parameters.

All of the above discussion has focused on measuring the black hole
mass and not the stellar orbital structure. The influence of these
effects on the orbits is harder to quantify, since the results depend
on which aspect of the orbits that concern us. For example, the answer
depends on whether one is concerned with the velocity ellipsoid at
every position in the galaxy, or whether one wants the radial to
tangential components at only two different radii.  The former is much
harder to measure. We are not at the point where we can study the
detailed shape of the velocity ellipsoid throughout the galaxy. The
two ingredients required to do this are (1) an understanding of any
systematic biases in the orbit-based techniques and (2) having the
appropriate data sets to perform this analysis. We will discuss each
of these below, but at this point we stress that obtaining a simple
measure of radial to tangential motion appears to be robust, and does
provide evolutionary constraints. Using the same Monte Carlo
simulations discussed above, we can also estimate the distribution of
radial to tangential motion from the noise in the spectra. The scatter
is remarkably small. Similarly, this ratio has very little dependence
on the smoothing parameter. In fact, that ratio changes by a much
smaller fraction than the best-fit black hole mass. This quantity is
typically measured to around 20\% or better. Thus, we are confident
that we can use this number to provide good comparison with
theoretical predictions.

\section{Results and Discussion}

There are 17 galaxies that have axisymmetric orbit-based
models. Figure~1.6 plots the orbital properties of those galaxies
against other galaxy properties. We include the black hole mass, the
effective dispersion, and the radial to tangential motion at two
points in the galaxy --- the central region and at 1/4 effective
radius. In the central region for each galaxy, the black hole
dominates the potential. The $M_\bullet-\sigma$ plot is the most
significant correlation. However, there is also a very strong
correlation between the black hole mass and the radial motion
contribution at large radii (top right plot). There is another
correlation of this quantity with effective $\sigma$ , but this may be
secondary to the one with black hole mass. In fact, the correlation
with black hole mass is the most significant of all other galaxy
properties (total light, total mass, effective radius, etc.). The
trend is that those galaxies that have large black hole masses (and
hence large $\sigma$) have orbits dominated by radial motion at large
radii. Tangential motion tends to occur in those galaxies with small
black holes. This correlation is one of the strongest for the full set
of comparisons in Gebhardt et al. (2003).


\begin{figure*}[t]
\centerline{\psfig{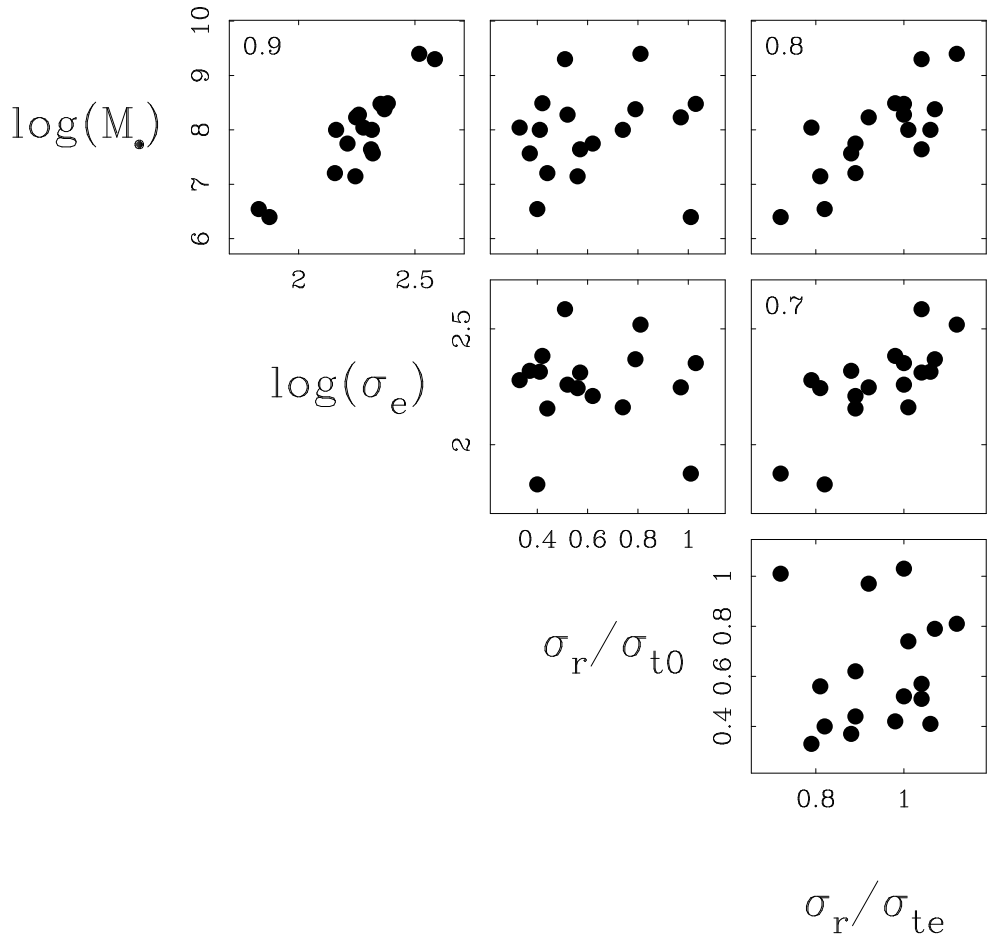}}
\vskip 0pt \caption{Plots of the orbital properties against various
galaxy properties. The properties along the diagonal include the black
hole mass, the effective dispersion ($\sigma_e$), the ratio of radial
to tangential dispersion at the center ($\sigma_r/\sigma_{t0}$), and
the ratio of radial to tangential dispersion at 1/4 the effective
radius ($\sigma_r/\sigma_{te}$). The number written in the upper left
corner of the plot is the Pearson's $R$ correlation coefficient. If
the probability from the correlation is below 10\%, we do not report
$R$.
\label{orbitc}}
\end{figure*}


The correlation between $M_\bullet$ and $\sigma_r/\sigma_{te}$ is
likely to be related to the evolutionary history of the galaxy. For
the most massive galaxies, at radii near to the effective radius, the
orbital distribution is radially biased. This is also the conclusion
from Cretton, Rix, \& de~Zeeuw (2000), who use orbit-based methods to
study the giant elliptical NGC~2320; along the major axis, they find
strong radial bias in the orbits at large radii. We can compare this
radial bias for the most massive galaxies with the $N$-body
simulations of Dubinski (1998). He finds that for the most massive
ellipticals, there is an increase in the radial motion from the center
(where it is nearly isotropic) to the outer radii (where the merger
remnant has $\sigma_r/\sigma_\theta = 1.3$). The most massive galaxies
in our sample of 17 approach this amount of radial motion at large
radii.  For the smaller galaxies, the $N$-body comparisons are not as
developed for measuring the internal orbital structure. However, based
on the recent results of $N$-body simulations (Meza et al. 2003;
Samland \& Gerhard 2003), we will soon be in a position to compare the
internal structure of the smaller galaxies as well. It has long been
known that low-luminosity ellipticals rotate rapidly and are often
consistent with oblate isotropic rotators, while high-luminosity
ellipticals have been thought to be supported by radial anisotropy at
large radii (Davies et al. 1983). Since black hole mass correlates
with luminosity, the $M_\bullet - \sigma_r/\sigma_{te}$ correlation
may then be secondary; however, the radial anisotropy correlates much
stronger with black hole mass than it does with luminosity. There has
been a considerable amount of theoretical work in explaining why the
black hole mass correlates so well with host galaxy dispersion (see
Adams et al. 2003 and references therein for a recent discussion).
The correlation may provide additional constraints on the models.

There is also a trend that the galaxies with shallow central density
profiles (i.e., the core galaxies) have orbits with the strongest
tangential bias near their centers. This correlation has been
discussed in Gebhardt et al. (2003). The most likely explanation is
that this is caused by binary black hole mergers. We know that the
existence of a black hole will leave some amount of tangential
anisotropy since it will either eject or accrete those stars that are
on radial orbits. This effect has been seen in many $N$-body
simulations that consider adiabatic growth of black holes (Quinlan et
al. 1995, 1997; Nakano \& Makino 1999; Milosavljevi\'c \& Merritt
2001; Sigurdsson 2003). In all of these case, however, the amount of
tangential motion is quite small. In the most detailed study to date,
Milosavljevi\'c \& Merritt (2001) find that the most extreme amount of
tangential motion has $\sigma_r/\sigma_\theta = 0.8$. The values that
Gebhardt et al. (2003) report are smaller than 0.4. One way to obtain
such large amounts of tangential motion is to have a binary black hole
that can affect more stars on radial orbits due to its own orbital
motion. The binary black hole results from a merger, and we already
have seen that binary black holes are one of the best mechanisms to
create the division between core and power-law galaxies (Faber et
al. 1997; Milosavljevi\'c \& Merritt 2001; Lauer 2003). However, the
$N$-body simulations that have been studied use fairly restrictive
assumptions --- most are based on spherical isotropic initial
conditions. Once realistic simulations including mergers and central
black holes are available, we will be in a much better position to
interpret the observational results.

\section{The Future}

There are many aspects of understanding the stellar orbital structure
that need improvement --- these include the data, analysis, and
theoretical comparisons. In regards to the data, with the use of
orbit-based models, we can realistically constrain the internal
structure of the galaxy. In fact, Verolme et al. (2002) were able to
measure with high accuracy the inclination of M32, and thus its
intrinsic shape. However, in order to do this they needed
two-dimensional kinematic data, which were obtained by the SAURON team
(de~Zeeuw et al. 2002). Most of the galaxies studied to date with
orbit-based models only have limited kinematic data (along 2--4
position angles) and thus cannot be used to study their intrinsic
shapes. In fact, as a result of this, most of the models in Gebhardt
et al. (2003) are only run as edge-on configurations, and there is a
concern that this may bias the results (de~Zeeuw 2003). However, for
the issues discussed here --- the black hole mass and radial to
tangential motion --- inclined models are unlikely to introduce
substantial changes, given the large uncertainties already on these
quantities.  In any event, significant improvement can be made by
using two-dimensional kinematic data. Another area for improvement of
the data is to include kinematics at large radii. In the study of
Gebhardt et al., they were careful to report only results inside of
the effective radii, where the dark halo is unlikely to have any
influence. However, any dynamical model needs to include some estimate
of the influence of orbits at large radii. Even though the effect of
these orbits is expected to be minimal at small radii, they are not
ignorable. In order to measure the central black hole and orbital
structure, a proper dynamical model should include both high-spatial
resolution (i.e., {\it HST}) and large-radii kinematics. With the
advent of integral-field units on many large ground-based telescopes,
obtaining this type of data will be feasible. In fact, adaptive optics
observations with an integral-field unit will be a tremendous advance
to this field of study.

On the data analysis side, while the orbit-based models that have been
run offer significant improvement over the previous set of models,
there is still a long way to go. For instance, most orbit-based models
are axisymmetric and oblate. Prolate and triaxial models need to be
included for a proper analysis. As discussed above, even for the
oblate models, most include only an edge-on configuration. In
addition, many have assumed luminosity density profiles that have
constant ellipticity with radius. We know that galaxies have
ellipticities and position angles that vary with radius, and so, at
some level, the models studied so far incorrectly represent the galaxy
light profile. However, at this point, the kinematic uncertainties
likely dominate the results, as opposed to assumption biases. One can
see this by comparing the inclined models for M32 (Verolme et
al. 2002) with the edge-on model of van~der~Marel et al. (1998). Even
there, the difference in the black hole mass is only at the 10\%
level, and the change in internal orbital structure is even
less. Since none of the other black holes are as well measured as
M32's (most have uncertainties around 30\%), this suggests that the
assumption biases will not have a great effect. Yet, once the quality
of the data improves, we will have to consider more general models. In
fact, triaxial models have already been studied by Verolme et
al. (2003). We know that kinematically distinct cores are common in
galaxies, and, therefore, axisymmetric models will clearly not provide
the best representation. Verolme et al. extend the orbit-based models
to include a triaxial distribution function and have successfully
reproduced the complicated kinematic structure of NGC~4365.  An
important step now would be to run both an axisymmetric and triaxial
model on the same galaxy to see if any significant differences arise.

The ultimate analysis method includes running an $N$-body model for
each galaxy. We know that at some level there is no galaxy that has
perfect symmetry. The question then becomes how significant are the
errors one makes when running a model that has some symmetry
(spherical, axisymmetric, or triaxial) to an asymmetric galaxy. At
least for the black hole mass, the errors are not large. Kormendy
(2003) summarizes the changes in black hole mass over time and with
different dynamical modeling sophistication. He finds that the change
in black hole mass, at least for a few well-studied galaxies, is not
very large, considering the enormous change in both data and
modeling. The black hole masses measured by Magorrian et al. (1998)
using low-quality ground-based data and 2-integral models measured
black hole masses to within a factor of 2--3 of the presently accepted
values. However, the intrinsic scatter of the $M_\bullet-\sigma$
correlation is consistent with zero, and at most 30\% (Tremaine et
al. 2002). Furthermore, the correlation of black hole mass with other
galaxy properties --- concentration index (Graham et al. 2001; Graham
2003) and total mass (Magorrian et al. 1998; McLure \& Dunlop 2002)
--- have a low scatter as well. The fact that the scatter in these
correlations is already so low implies that the systematic
uncertainties are not terribly measured; otherwise, we would not be
able to detect these correlations. In order to better study these
correlations, we must have better determined black hole masses, and
therefore we must improve the analysis techniques. Hopefully, we will
not have to measure black hole masses to much better than 10\% to
answer the scientifically important questions, since going beyond that
will be a challenge in terms of both observations and analysis.

\end{document}